\documentclass[twocolumn, nofootinbib, amsmath, amssymb, citeautoscript, superscriptaddress]{revtex4-2}

\usepackage{graphicx}
\usepackage{dcolumn}
\usepackage{bm}
\usepackage{braket}
\usepackage[%
colorlinks=true,
urlcolor=blue,
linkcolor=blue,
citecolor=blue
]{hyperref}
\usepackage{cleveref}
\usepackage{verbatim}
\usepackage{xcolor}
\usepackage{multirow}
\usepackage{physics}
\usepackage{mathtools}
\usepackage{bbold}
\usepackage{comment}
\usepackage{placeins}
\usepackage{enumitem}
\usepackage{tabularx}
\usepackage{cellspace} % New package for cell padding
\usepackage{dsfont} % for the symbol of Z2 index

\newcommand{\beq}{\begin{equation}} 
\newcommand{\eeq}{\end{equation}}
\def\mathclap#1{\text{\hbox to 8pt{\hss$\mathsurround=0pt#1$\hss}}}
\newcommand{\cc}{}

\begin{document}
	
%	\setlength{\abovedisplayskip}{7.9pt}
%	\setlength{\belowdisplayskip}{7.9pt} 

%\preprint{APS/123-QED}

\title{First-principles studies of fermiology in topological phases of bulk ZrTe$_5$}

\author{Chao Chen Ye}\thanks{c.chen.ye@rug.nl}
\author{Yuliia Kreminska}
\author{Jianting Ye}
\author{Jagoda Sławińska}\thanks{jagoda.slawinska@rug.nl }
\affiliation{Zernike Institute for Advanced Materials, University of Groningen, Nijenborgh 3, 9747 AG Groningen, Netherlands}

%\date{\today}

\begin{abstract}
Topological insulators have been studied intensively over the last decades. Among these materials, three-dimensional (3D) zirconium pentatelluride (ZrTe$_5$) stands out as one of the most intriguing for both theoretical and experimental studies because of its diverse range of distinct topological phases. In this work, we employ density functional theory to study the electronic structure and quantum oscillations exhibited by various topological phases of 3D bulk ZrTe$_5$. We have discovered that by analyzing combined patterns in band structures, Fermi surfaces, and Shubnikov-de Haas (SdH) oscillations we can determine the corresponding topological phase without relying on the conventional calculation of topological invariants or boundary state contributions. This approach facilitates the identification of topological phases in  ZrTe$_5$ directly from experimental quantum oscillation measurements. Using this method, we have analyzed the entire process of topological phase transition, revealing changes in the topology of the Fermi pockets and validating the shapes deduced from the experimental data for the topological phases.
\end{abstract} 	

%\begin{abstract}
%	Topological insulators have been studied intensively over the last decades. Three-dimensional (3D) zirconium pentatelluride (ZrTe$_5$) is one of the most intriguing materials for both theoretical and experimental studies because of the richness of distinct topological phases. In this work, we employ density functional theory to study electronic structure and quantum oscillations of different topological phases of 3D bulk ZrTe$_5$. We have unveiled patterns in band structures, isoenergetic surfaces in the reciprocal space (i.e. isosurfaces), and Shubnikov-de Haas (SdH) oscillations that allow one to determine the corresponding topological phase without calculating any topological invariant or boundary state contributions, which greatly facilitates the interpretation of theoretical and experimental results for this material. We have also analyzed the entire process of topological phase transition in both band structures and isosurfaces. Furthermore, our computed isosurfaces validate the surface shape predictions deduced from the experimental data. Finally, we have identified that variations in a specific lattice parameter are the key term which leads to the bulk gap closing, something that was not clear in prior theoretical and experimental studies.
%\end{abstract} 	

\maketitle 

%%%%%%%%%%%%%%%%%%%%%%%%%%%%%%%%%%%%%%%%%%%%%%%%%%%%%%%%%%%%%%%%%%%
\section{Introduction}
\label{sec:introduction}

Topological insulators (TIs) are materials that possess simultaneously an insulating bulk and conducting boundary states which are intrinsically protected by symmetries, making them resilient against perturbations~\cite{hasan_2010, qi_2011, asboth_book}. In analogy to distinct phases of matter (\textit{e.g.}, liquid, solid, and gas), TIs present various topological phases~\cite{asboth_book, hasan_2010}. Although there are abundant theoretical studies of TIs thus far~\cite{ fu_2007, asboth_book, hasan_2010,  schnyder_2008, bergholtz_2020}, a limited number of materials are suitable for experimental studies of distinct topological phases via direct Fermi surface measurement by quantum oscillations. This is because such studies require high-mobility materials with pronounced oscillations in transport measurements. A perfect example is the three-dimensional (3D) zirconium pentatelluride (ZrTe$_5$)~\cite{fjellvag_1986, weng_2014, manzoni_2016, wu_2016, zhang_2017, fan_2017, tang_2019, zhuo_2022, wieder_2022, facio_2023, kovacs_2023}, which has two topological phases named strong TI (STI) and weak TI (WTI) that transform seamlessly by bridging with a Dirac semimetal, having a band-gap closing~\cite{hasan_2010, qi_2011, weng_2014, manzoni_2016, fan_2017, zhang_2021, facio_2023}. With the rich topological behavior, experimentally, ZrTe$_5$ exhibits numerous unusual quantum phenomena, including the nonlinear anomalous Hall effect, chiral magnetic effect, 3D quantum Hall effect (QHE)~\cite{tang_2019, galeski_2021}, log-$B$ quantum oscillations~\cite{wang_2018}, and superconductivity~\cite{zhou_2016}. All these effects make ZrTe$_5$ intriguing to explore both theoretically and experimentally. 

%These phenomena are generally ascribed to the presence of massive Dirac fermions near the energy band closing point because these electronic states have large velocities, long wavelengths, and are robust against backscattering~\cite{zhuo_2022, facio_2023}.

Current experimental evidence strongly suggests that most of the stable 3D ZrTe$_5$ samples are in the WTI phase ~\cite{tang_2019, galeski_2021, facio_2023, kovacs_2023}. However, density functional theory (DFT) simulations, using the experimentally obtained lattice parameters, predict the STI phase ~\cite{fjellvag_1986, fan_2017, kovacs_2023}. As ZrTe$_5$ resides near the WTI-STI transition, the exact topological phase is notably sensitive to subtle variations in the structural parameters of the material~\cite{facio_2023}. This sensitivity suggests that external strains that change lattice parameters and atomic positions can serve as a knob for inducing topological phase transitions~\cite{manzoni_2016, weng_2014, fan_2017, facio_2023}. In theoretical studies, one can simulate this tuning process by varying lattice parameters to form different topological phases, hence exploring their properties. 

%Finally, besides its easy-tuning characteristics of topological phases, ZrTe$_5$ also exhibits numerous unusual quantum phenomena, including the anomalous Hall effect, chiral magnetic effect, 3D quantum Hall effect (QHE), and log-$B$ quantum oscillations. These phenomena are generally ascribed to the presence of massive Dirac fermions~\cite{zhuo_2022}.

The experimental identification of different topological phases for a specific ZrTe$_5$ sample is highly nontrivial. One effective technique involves the analysis of quantum oscillations which can yield the Berry phase in a Landau index plot~\cite{zhu_2022, kovacs_2023}. Quantum oscillations also provide the information about the extremal orbits of isoenergetic surfaces in the reciprocal space, commonly known as isosurfaces or  Fermi surfaces~\cite{landau_2013, kamm_1985} which are sensitive to the change of lattice parameters or carrier doping in doped semiconductors. Thus, these measurements allow the identification of potential Dirac cones. Particularly, for 3D ZrTe$_5$,  quantum oscillation measurements close to the band-gap closing point can potentially unveil a unique bulk Dirac cone with a presumed small 3D ellipsoidal isosurface~\cite{tang_2019, galeski_2021}. The Dirac fermions are claimed to be the origin of the measured 3D QHE ~\cite{tang_2019}, according to Halperin's predictions~\cite{halperin_1987}. On the other hand, there are surprisingly no theoretical studies of isosurfaces to be compared with these experimental reports of the associated quantum oscillations. Therefore, several fundamental points such as the number of Fermi pockets at the Fermi level remains unclear~\cite{facio_2023}. It is important when one wants to study Dirac fermions, for instance, to confirm the existence and origin of the 3D QHE ~\cite{tang_2019}.

%conflicting conclusions arise from different quantum oscillation measurements and \textit{ab-initio} studies regarding the existence or absence of additional Fermi pockets close to the Fermi level, which is vital to identify possible Dirac cones~\cite{facio_2023}.

In this paper, we perform DFT simulations to identify different topological phases and analyze topological phase transitions of the 3D bulk ZrTe$_5$. We start by introducing the properties of the material and computational methodology in Sec.~\ref{sec:material_and_method}. In Sec.~\ref{sec:results}, we present the simulation results and discuss their impact on the field. Sec~\ref{subsec:bands} shows the energy band structure when continuously deforming the atomic structure of the material. These results allow us to identify the entire process of topological phase transition. Isosurfaces shown in Sec~\ref{subsec:fs} unveil a clear distinction between different phases of bulk ZrTe$_5$. They can be used directly to determine the corresponding topological phase of ZrTe$_5$, thereby helping to validate the isosurfaces from experimental outcomes~\cite{tang_2019, galeski_2021}. Finally, Sec.~\ref{subsec:qo} presents the calculation of the Shubnikov-de Haas effect, an archetypal quantum oscillations experiment conducted for quantitatively describing the isosurfaces. The comparison of our calculated results with the experimental data reveals that ZrTe$_5$ samples are indeed sensitive to the structural parameters. This is consistent with the widely diverse physical properties reported on this material as the actual physical properties are highly dependent on crystal fabrication and device measurement conditions that can cause subtle changes in doping and crystal lattice parameters. For ZrTe$_5$, in general, we have unveiled insights that allow one, by simultaneously linking theoretical simulation with the transport experiments, to determine the corresponding topological phase of the material, without referring to any specific calculation about topological invariants or boundary state contributions.

\section{Material and methodology}
\label{sec:material_and_method}
\subsection{ZrTe$_5$}
\label{subsec:material}

\begin{figure*}[htb!]
	\centering
	\includegraphics[width=0.3\linewidth]{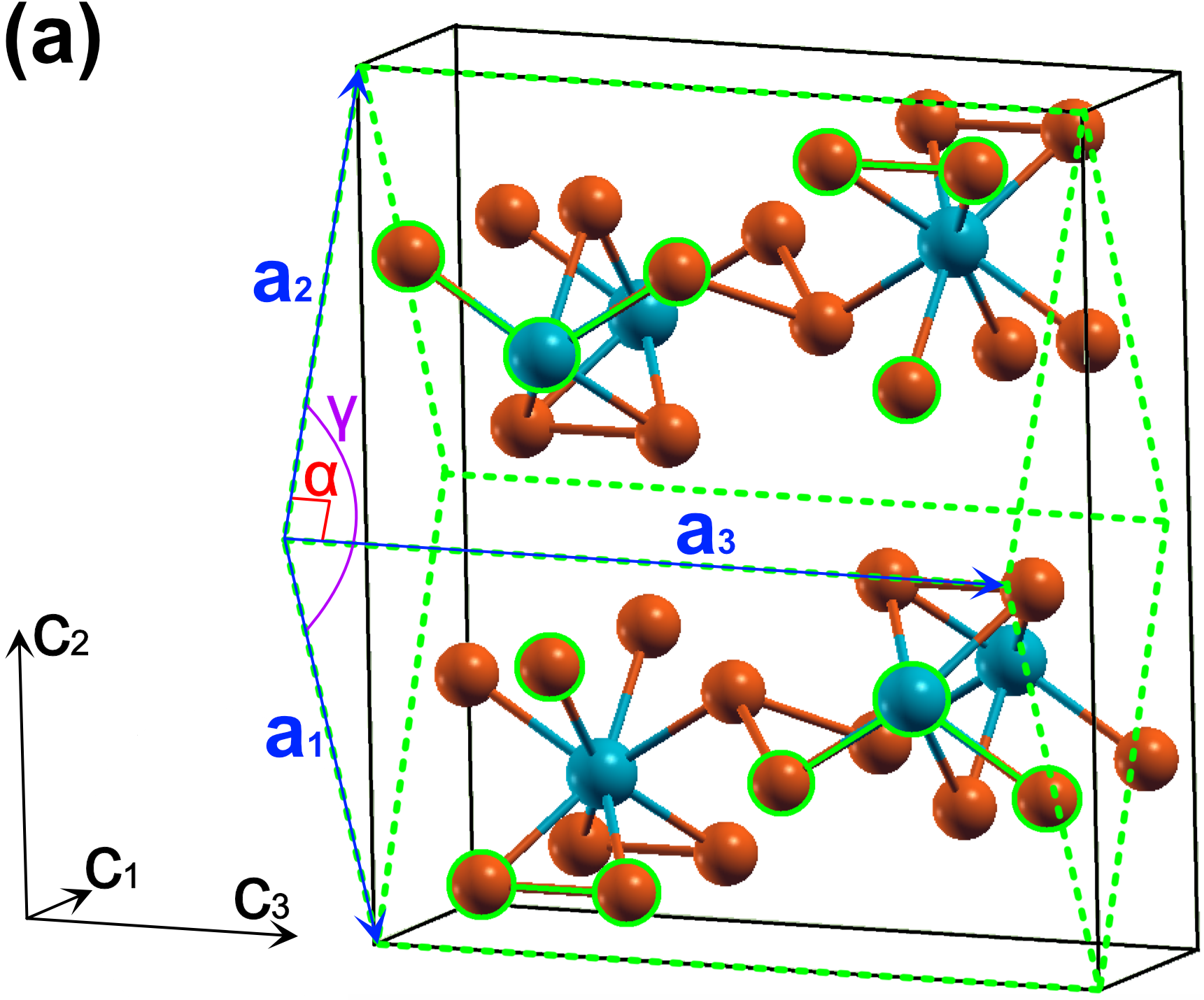}
	\includegraphics[width=0.36\linewidth, height=6cm, keepaspectratio]{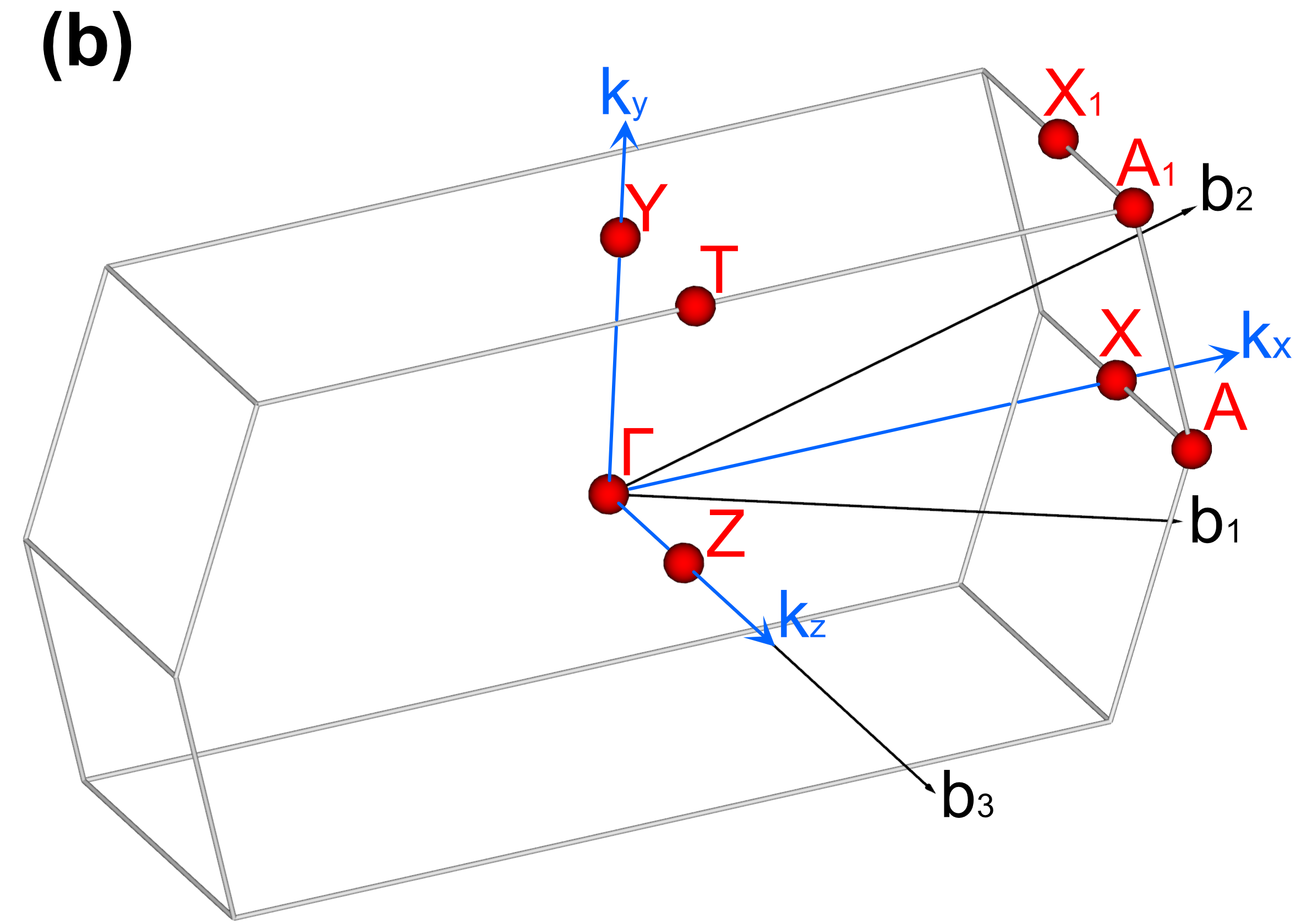}
	\includegraphics[width=0.3\linewidth, height=6cm, keepaspectratio]{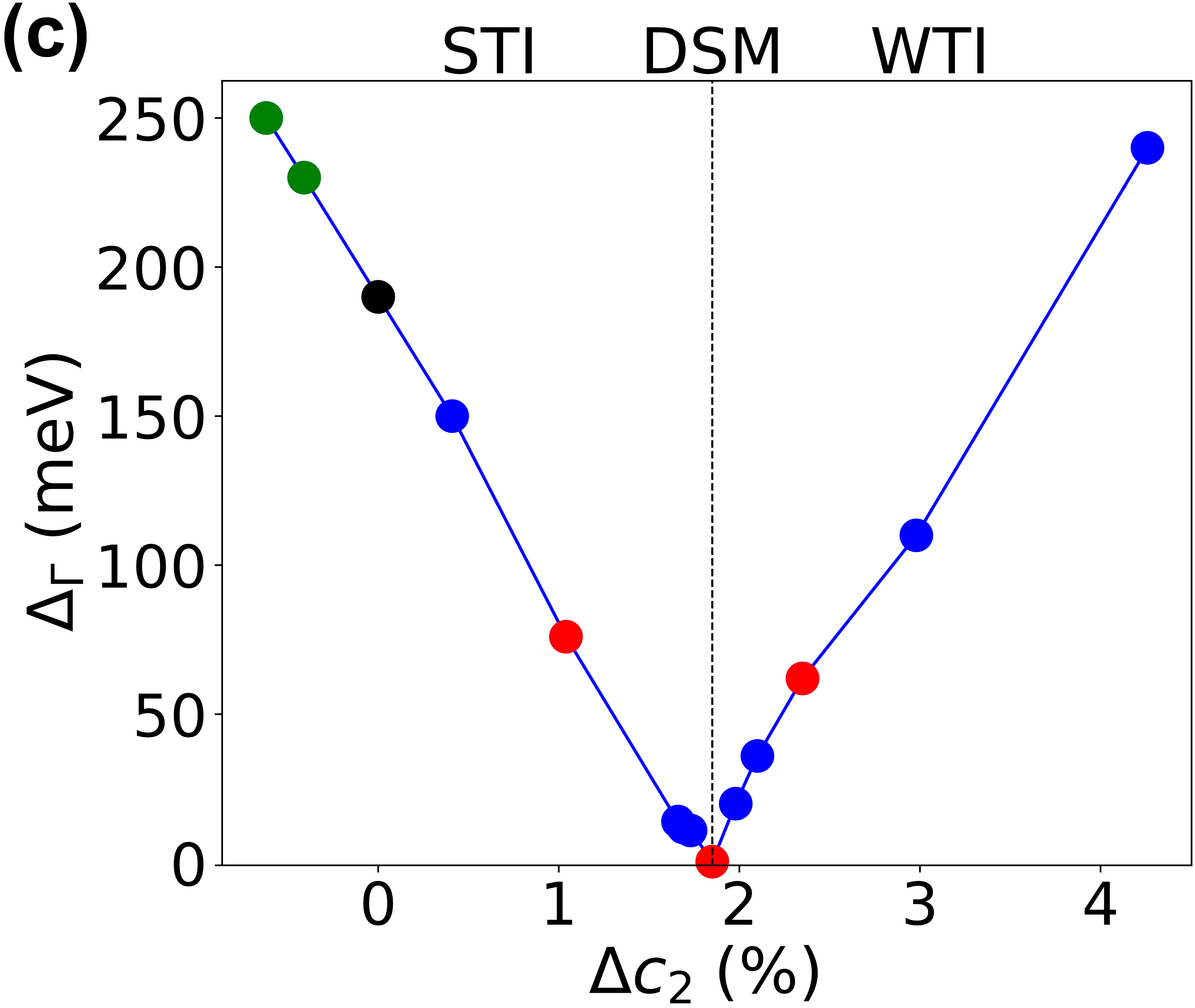}
	\caption{(\textbf{a}) Crystal structure of 3D ZrTe$_5$ in the conventional unit cell (black solid lines with all ions) and primitive cell (green dashed lines with green-color wrapped ions). $\qty{\vb{a}_i}_1^3$ are the lattice vectors of the primitive cell, and $\qty{\vb{c}_i}_1^3$ are the scaled ones of the conventional cell. (\textbf{b}) The first Brillouin zone for the primitive cell with the high-symmetric points. (\textbf{c}) Evolution of the bulk band gap at $\Gamma$ point when the $c_2$-axis changes; the continuous evolution is the result of small and constant structural deformations. The phase transition between STI and WTI is highlighted by the continuous evolution of $\Delta c_2$ inside a topological phase upon the gap closing at DSM when $\Delta c_2 = 1.85 \; \%$. The green dots correspond to experimental lattice parameters; the black dot corresponds to the DFT fully-relaxed system. The red dots are the configurations chosen for further analysis.}
	\label{fig:crystal_structure}
\end{figure*}

ZrTe$_5$ possesses a base-centered orthorhombic crystal structure with \textit{Cmcm} (No. 63) space group symmetry~\cite{weng_2014, fan_2017}. The 3D structure comprises a bilayer system in the $c_1-c_3$ plane, stacked along the $c_2$-axis via weak van der Waals interactions. This arrangement forms a conventional unit cell containing two Zr and ten Te atoms per layer (Fig.~\ref{fig:crystal_structure}\textit{a}, black contour). The primitive cell, however, assumes a rhombus shape with depth (\textit{i}.\textit{e}. a parallelepiped), and includes half the number of atoms that are contained in the conventional cell (Fig.~\ref{fig:crystal_structure}\textit{a}, green contour). In our DFT calculations, we employ the primitive cell \cite{weng_2014, fan_2017}. The corresponding Brillouin zone (BZ) with high-symmetry points is illustrated in Fig.~\ref{fig:crystal_structure}\textit{b}. 

The STI is characterized by a bulk energy gap, accompanied by gap-closing boundary states on each 2D surfaces. The coupling of these 2D boundary states forms a phase resembling 3D quantum Hall state, resulting in a single 3D surface Dirac cone in the energy band structure~\cite{kohmoto_1992, fu_2007, ringel_2012, manzoni_2016, fan_2017}. These topologically distinctive boundary states, existing on all surfaces, are robust against disorder and external perturbations~\cite{fu_2007, ringel_2012}. In contrast, the WTI also has a bulk energy gap, but its boundary states are restricted to specific high-symmetry surfaces due to the inherent material anisotropies~\cite{ringel_2012}. Consequently, these states do not form any surface Dirac cone and can be more susceptible to perturbations such as disorders and surface terminations. For instance, they can become gapped without breaking the time-reversal symmetry or causing any closing of the bulk gap; this effect occurs when disorder breaks the translational symmetry along $c_2$-axis~\cite{fu_2007, ringel_2012}.

The topological phase transition between STI and WTI phases is manifested by a closure of the direct bulk energy gap at high-symmetry $\Gamma$ point~\cite{weng_2014, manzoni_2016, fan_2017, tang_2019}. This leads to the formation of a bulk Dirac cone preserved by the inversion and time-reversal symmetries, resulting in the characteristics of a Dirac semi-metal (DSM)~\cite{weng_2014, fan_2017}. It is important to note that while DSM is not classified as a topological phase, it does represent a phase characterized by Dirac physics near the Fermi level. This phase is particularly intriguing due to the unique properties of Dirac fermions~\cite{chen_chen_2015, chen_2015, yuan_2016, liu_2016, pariari_2017,  tang_2019, konstantinova_2020}. Finally, the stable crystal structure of 3D ZrTe$_5$ is located near DSM, allowing easy exploration of different phases. 

\subsection{Computational method}
\label{subsec:method}

We used DFT implemented in the \texttt{QUANTUM ESPRESSO} package~\cite{qe_2009, qe_2017} to investigate the ground states of 3D bulk ZrTe$_5$ under various conditions that deform the material at the atomic level; the method discerns the potential topological phase transitions~\cite{manzoni_2016, fan_2017, xu2018temperature}. We manually changed the lattice parameters of the primitive cell, where the variations can simulate the structural deformation that can occur due to, for instance, the presence of applied external strain~\cite{manzoni_2016, fan_2017, xu2018temperature}. For simplicity, we refer to each of these structures as \textit{Config.} hereafter.

For this study, we adopted the generalized gradient approximation (GGA) in the form of the Perdew, Burke, and Ernzerhoff functional~\cite{pbe_1996}. The ion-electron interactions were treated using fully relativistic projector-augmented wave pseudopotentials taken from the \texttt{PSLIBRARY} database~\cite{pp_2014}. The plane-wave cutoff energy was set to $80 \,$ Ry, and the $k$-point mesh was $16 \times 16 \times 8$ for the self-consistent computations and $30 \times 30 \times 10$ for the non-self-consistent ones. Spin-orbit coupling (SOC) was taken into account in all the calculations, except during atomic structural optimizations. The wave functions projected from \texttt{QUANTUM ESPRESSO} onto pseudo-atomic orbitals served to construct tight-binding (TB) Hamiltonians. The latter was done using the post-processing tool \texttt{PAOFLOW}~\cite{paoflow_2018, paoflow_2021}. The Hamiltonians were interpolated on denser \(k\)-grids of $100 \times 100 \times 18$ to generate accurate and smooth isosurfaces.

We further used the \textit{Supercell K-space Extremal Area Finder} (\texttt{SKEAF})~\cite{skeaf_2012} to calculate the Shubnikov-de Haas (SdH) effect for various isosurfaces close to the valence band maximum (VBM); this effect is a type of quantum oscillations where the electrical resistivity (conductivity) oscillates~\cite{kittel_2004}. To be specific, we calculated SdH oscillation frequencies from the Onsager relation~\cite{kittel_2004, tang_2019},

\begin{equation}
	B_F = \frac{1}{\Delta \qty(1/B)} = \frac{\hbar}{2 \pi e} A_{extreme} \; ,
	\label{eq:SdH_frequencies}
\end{equation}
and the corresponding cyclotron mass~\cite{kittel_2004},
\begin{equation}
	m_{CR}= \frac{\hbar^2}{2 \pi } \pdv{A(E, k_\parallel)}{E} \; .
	\label{eq:cyclotron_mass}
\end{equation}

Here, $e$ is the elementary charge, $A$ indicates the cross-sectional area of the isosurface with energy $E$ in a plane perpendicular to the magnetic field vector $\vb{B}$, $A_{extreme}$, referring to the corresponding extremal cross-sectional areas, and $k_\parallel$ is the component of the momentum vector $\vb{k}$ parallel to $\vb{B}$. Note that, according to Eqs.~\eqref{eq:SdH_frequencies} and \eqref{eq:cyclotron_mass}, the quality and smoothness of the calculated isosurfaces are essential for accurately determining the SdH oscillation frequencies and cyclotron masses.

% the SdH frequencies $B_F$, Eq.~\eqref{eq:SdH_frequencies}, and the corresponding cyclotron mass $m_{CR}$, Eq.~\eqref{eq:cyclotron_mass}, for various isosurfaces close to the Fermi energy in the valence band. 

\section{Results and discussion}
\label{sec:results}

%%%%%%%%%%%%%%%%%%%%%%%%%%%%%%%%%%%%%%%%%%%%%%%%%%%%%%%%%%%%%%%%%%%%
\subsection{Band structure}
\label{subsec:bands}

\begin{figure*}[tb!]
	\centering
	\includegraphics[width=0.8\linewidth]{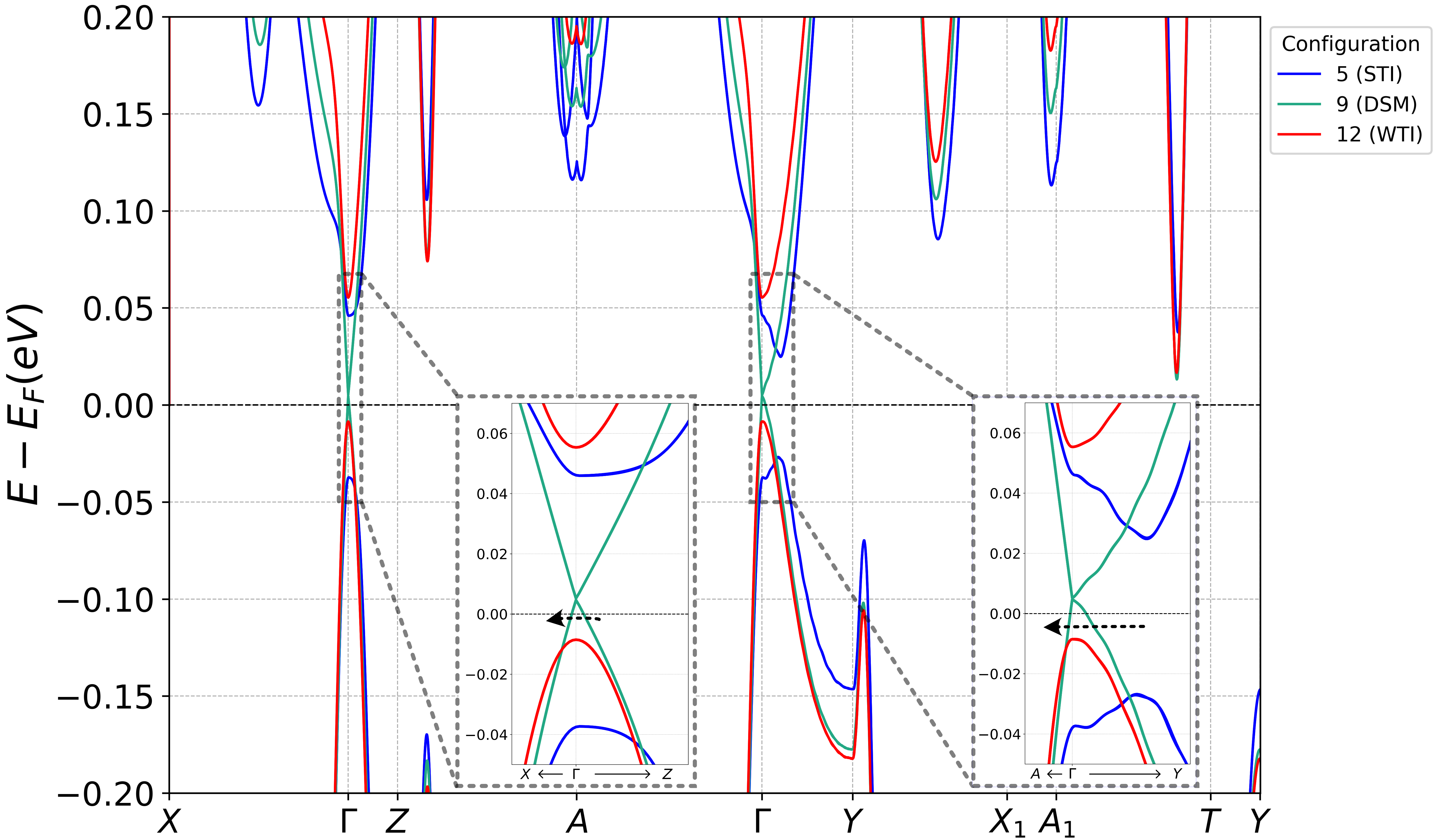}
	\caption{Evolution of the bulk band structure of ZrTe$_5$ for the three chosen structures marked as red dots in Fig.~1; they are \textit{Configs.} 5, 9, 12 listed in SM\cc, representing STI, DSM, and WTI, respectively. The band energy difference close to $\Gamma$ point is expanded to show the details in the insets, where the dashed arrow tracks the shift of the valence band maximum.}
	\label{fig:bands_reduced}
\end{figure*}

We start by analyzing the difference in the energy band structure of all \textit{Configs.}, which is achieved by tuning distinct lattice parameters as detailed in the Supplementary Material (SM)\cc. We use the high-symmetry points illustrated in Fig.~\ref{fig:crystal_structure}\textit{b}; the path $X-\Gamma-Z-A-\Gamma-Y-X_1-A_1-T-Y $, chosen to represent the band structures, contains all relevant information. Again, it is essential to note that our DFT simulations focus on bulk states, and do not include boundary contributions. 

Structural deformations lead to a continuous evolution of the gap closing at the $\Gamma$ point, $\Delta_\Gamma$~\cite{weng_2014, fan_2017, tang_2019, zhu_2022}, as shown in Fig.~\ref{fig:crystal_structure}\textit{c}. Here, we represent the evolution in terms of relative changes on $\vb{c}_2$-axis of the conventional cell because lattice parameters of the primitive cell have two symmetric components,  $\vb{a}_1$ and $ \vb{a}_2$ which do not directly determine topological phase transition. The reason for choosing $\Delta  c_2 \cdot \vb{\hat{c}}_2= \Delta \qty(\vb{a}_1 + \vb{a}_2 ) = 2 \,  \Delta a_1 \cdot \vb{\hat{c}}_2$ is due to the continuous evolution of $\Delta_\Gamma$ without sudden changes within a topological phase, as depicted in Fig.~\ref{fig:crystal_structure}\textit{c}. This is the consequence of the adiabatic theory for topological materials~\cite{asboth_book, zhu_2022}; other components do not show this behavior, as shown in SM \cc (Fig. S1). With the increase of $\Delta  c_2$, the gap, $\Delta_\Gamma$, linearly decreases to zero and subsequently increases, indicating a topological phase transition~\cite{asboth_book, fan_2017}. The fact that $\Delta  c_2$ is the key term that determines the gap closing was not evident in prior theoretical and experimental studies that focused on changes in volume or interlayer distance (see SM for details\cc)~\cite{weng_2014, manzoni_2016, fan_2017}.

%Since the gap closing happens exactly at $\Gamma$ point~\cite{weng_2014, fan_2017, tang_2019}, $\Delta_\Gamma$, this quantity is listed in Tab.~\ref{tab:cell_parameters} together with relative changes of the primitive cell volume, $\Delta V$, and conventional cell parameter $a_2^\text{conv}$, $\Delta a_2^\text{conv} = \Delta \qty(a_1 + a_2 ) = 2 \Delta a_1 $, for each \textit{Config.} As we progress from \textit{Configs.} 1 to 14, it is evident that the gap linearly decreases until nearly zero in \textit{Config.} 9\footnote{Achieving the exact zero in DFT simulations is challenging due to \textit{(i)} the difficulty in finding precise lattice parameters and \textit{(ii)} the value $\Delta_\Gamma = 0.55 \, \text{meV}$ being within DFT errors, making it uncertain whether it represents a real gap or an artefact.}, subsequently, it starts increasing. This phenomenon gives sign of a topological phase transition~\cite{asboth_book, fan_2017}. Notably, the linear variation is observed in $\Delta a_2^\text{conv}$ rather than $\Delta V$; it means that $a_2^\text{conv}$, which is along the stacking axis of monolayers, is the key component that affect the gap closing. This insight was not clear in prior theoretical and experimental studies~\cite{weng_2014, fan_2017, manzoni_2016}. Fig.~\ref{fig:crystal_structure}\textit{c} shows this evolution.

The black dot in Fig.~\ref{fig:crystal_structure}\textit{c} is our fully-relaxed system without introducing any deformation, where the lattice parameters coincide with those reported in Ref.~\cite{fan_2017} when considering three decimals in accuracy (see SM for details\cc). Notably, results may vary among different groups in both computational and experimental studies due to distinct approaches employed. Nevertheless, the underlying physics remains consistent for the correspondence between the topological phase and the specific \textit{Config.}~\cite{weng_2014, fan_2017, kovacs_2023}. Consequently, it can be asserted that the \textit{Config.} represented by the black dot corresponds to STI~\cite{fan_2017}. The green dots in Fig.~\ref{fig:crystal_structure}\textit{c} are calculated by adopting the measured lattice parameters from Refs.~\cite{tang_2019, fjellvag_1986}. The evolution of the band gap at $\Gamma$ point implies that the above three \textit{Configs.} (\textit{i}.\textit{e}., green and black dots) all belong to the same topological phase (\textit{i}.\textit{e}., STI phase), according to the adiabatic theory. This is consistent with the reported results~\cite{weng_2014, fan_2017, kovacs_2023}.

Figure~\ref{fig:bands_reduced} shows representative band structures for different $\Delta c_2 \; \qty{< 1.85\%, = 1.85 \%, > 1.85\%} $ according to Fig.~\ref{fig:crystal_structure}\textit{c}. The chosen \textit{Configs.} for each case are represented by red dots, which belong to \textit{Configs.} 5, 9, 12 (see SM for details\cc). To compare different \textit{Configs.}, we consider the reference as $E-E_F$ from our DFT calculations; however, only the Fermi level of DSM, $E_F = 6.381$ eV, has a physical meaning because it truly has occupied states at  $E = E_F $, as one can see in the band structure. The insets highlight the energy band difference close to $\Gamma$ point, where the dashed arrow tracks the shift of VBM. Focusing on the $\Gamma$ point, VBM of the \textit{Config.} 5 is located along the $Z -\Gamma - Z$ and $Y -\Gamma - Y$ paths, whereas \textit{Config.} 12 has it along the $X - \Gamma -X$ and $A - \Gamma -A$ paths (see SM for a more detailed band structure). The transition of VBM occurs at \textit{Config.} 9, where a gap closing appears. Physically, different high-symmetric points usually do not have the same symmetries, hence this transition reflects that symmetry changes in the system upon topological phase transition. We have deduced previously that \textit{Configs.} 5 and 9 belong to STI and DSM, respectively. Therefore, \textit{Config.} 12 corresponds to WTI. Furthermore, this identification of topological information implies that distinction among intermediate \textit{Configs.} results from adiabatic deformations of the system within the respective topological phases. Thus, we can keep tracking the entire topological transition, as marked by the black dashed arrow in the insets of Fig.~\ref{fig:bands_reduced}. Moreover, this feature leads to a pattern in the band structure that tells us the corresponding topological phase, without calculating any topological invariant or boundary state contributions.

Associated with the topological change, an additional concern is the number of Fermi pockets, which becomes relevant when the material is doped. It is worth noting that the number of pockets reported differs widely among various authors~\cite{fan_2017, facio_2023}. Figure~\ref{fig:bands_reduced} shows that the conduction band minimum never appears near the $\Gamma$ point, except for the DSM phase (see other band structures in SM\cc, Fig. S2). Thus, one can find more than one Fermi pocket with different topologies. In the case of DSM, it has pure Dirac physics for energies below $E-E_F=13.1$ meV. This result is also consistent with experimental measurements in Ref.~\cite{tang_2019}, despite a slight difference in value. The continuous evolution implies that a system must be very close to DSM to have a single Fermi pocket in the conduction band located at the $\Gamma$ point. In contrast, VBM generally appears near the $\Gamma$ point, and the rest of the band edges lie much lower in energy. Therefore, even with large doping, state could still be located close to the $\Gamma$ point. Furthermore, the VBM in the right inset generally lies along $Y - \Gamma - Y$ path for all deformations in STI, resulting in a pair of symmetric Fermi pockets. In the case of WTI, the Gaussian-type maximum typically implies a single Fermi pocket, except in the extreme WTI case such as \textit{Config.} 14 (see SM for details\cc). 

\subsection{Isosurfaces}
\label{subsec:fs}
While band structures along the high-symmetry lines offer valuable insights, the electronic structure across the whole BZ is essential to fully understand the topology. Isosurfaces provide more information about the system and allow the comparison of theoretical and experimental quantum oscillation results. In this context, we present the isosurfaces at different energies (doping) in the valence band for different topological phases. Focusing on Fermi pockets close to $\Gamma$ point, we continue with exemplary \textit{Configs.} $5, 9, 12$ representing STI, DSM and WTI phases, respectively. Since the physics is the same within a topological phase according to the adiabatic theory, we will primarily study these \textit{Configs.} henceforth. Thus, they are conveniently referred to by their corresponding topological phase for simplicity.

%This choice relies on the similarity in terms of band gap at $\Gamma$ point and the proximity of $\Delta a_2^\text{conv} $ respect to the topological phase transition point, as shown in Tab.~\ref{tab:cell_parameters} and by red dots in Fig.~\ref{fig:crystal_structure}. Furthermore, since the physics is the same inside a topological phase according to the adiabatic theory, we will primary study these \textit{Configs.} henceforth. Thus, they are conveniently referred to by their corresponding topological phase for simplicity.

\begin{figure*}
	\centering
	\includegraphics[width=\linewidth]{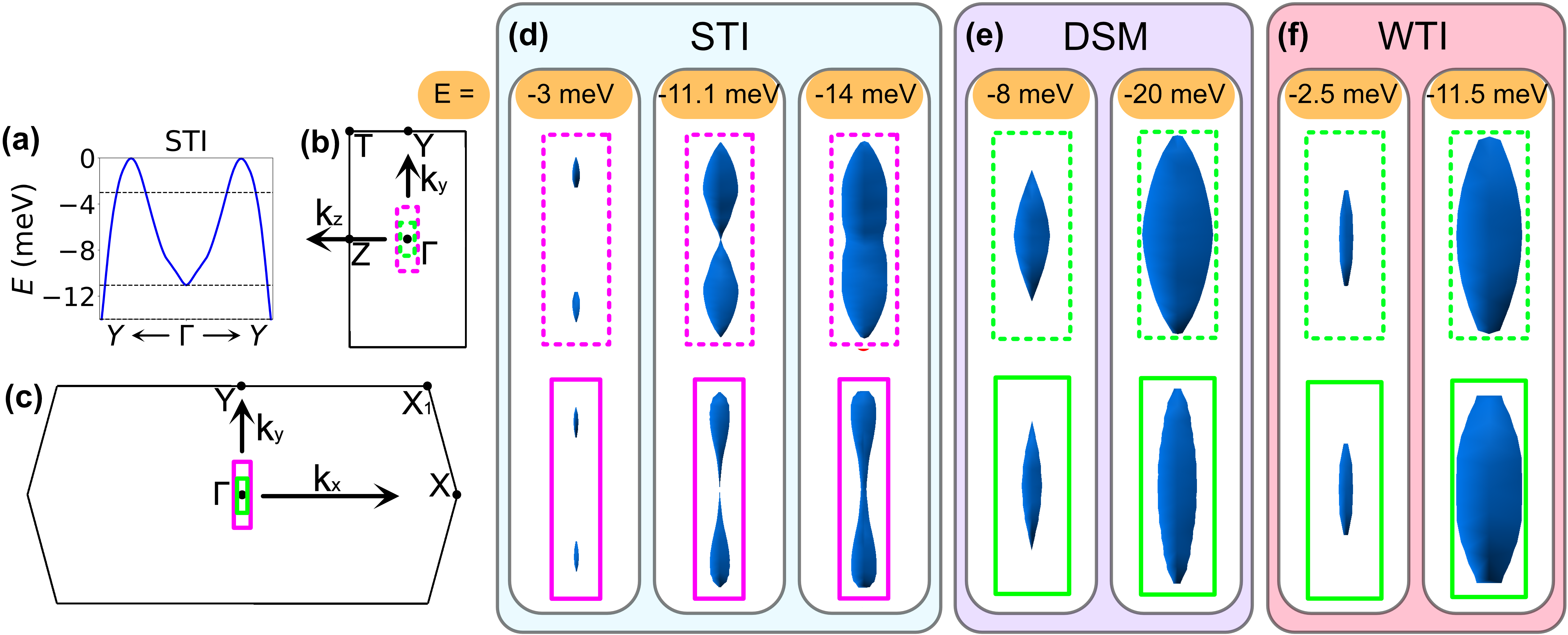}
	\caption{(\textbf{a}) Band structure of STI along $Y - \Gamma - Y$ path that gives insight on the evolution from the symmetric Fermi pockets into a single one. The dashed lines are the isoenergies to represent the isosurfaces for STI. (\textbf{b}), (\textbf{c}) Tiny areas where the isosurfaces appear in the Brillouin zone, viewed from the $k_x$- and $k_z$-axis, respectively. Evolution of the valence isosurfaces for different values of chemical potential of (\textbf{d}) STI (\textit{i.e.}, \textit{Config.} $5$), (\textbf{e}) DSM (\textit{i.e.}, \textit{Config.} $9$) and (\textbf{f}) WTI (\textit{i.e.}, \textit{Config.} $12$). We set the energy references $E_{VBM}=0$ at the VBM of each configuration.} 
	\label{fig:fs_main_text}
\end{figure*}

Figure ~\ref{fig:fs_main_text} illustrates the phase transition and the topology of the Fermi pockets in each phase. Close to VBM, all isosurfaces are located in tiny regions of the Brillouin zone centered at $\Gamma$ point. Figure ~\ref{fig:fs_main_text}\textit{b, c} highlight the areas by purple and green rectangles from the $k_z$ and $k_x$-axis viewpoint, respectively. Starting with the evolution of isosurfaces by decreasing isoenergies, Fig.~\ref{fig:fs_main_text}\textit{d} shows that the STI initially has a pair of symmetric Fermi pockets along $k_y$-axis for $E = -3$ meV as discussed in the previous section. Upon moving down in energy from VBM, the pockets increase their size and merge into a single one at $E = -11.1$ meV, with a narrow connection at $\Gamma$ point. As the downshift in energy continues, the thickness of the connection point enlarges quickly, especially in the $(k_y, k_z)$-plane, as shown for $E = -14$ meV. Finally, it results in a non-uniform elongated 3D ellipsoid, which is wider when viewed from $k_x$ (\textit{i}.\textit{e}., top figures) than $k_z$-axis (\textit{i}.\textit{e}., bottom figures). This evolution can be tracked from the band structure in Fig.~\ref{fig:fs_main_text}\textit{a}, where we see the two symmetric peaks along the $Y - \Gamma - Y$ path. In the case of DSM, the non-uniform Fermi pocket always appears as an elongated 3D ellipsoidal, like the final stage of the STI phase. The decreasing isoenergies only exhibit as a simple increase of the volume by keeping in overall shape, as shown in Fig.~\ref{fig:fs_main_text}e. The evolution of the Fermi pocket of the WTI is similar to the DSM case, but the elongated 3D ellipsoidal is more uniform (see Fig.~\ref{fig:fs_main_text}\textit{f}). The non-uniform elongated 3D ellipsoidal shape of our calculated Fermi pockets of DSM and WTI proves the predicted outcome from experimental results in Refs.~\cite{tang_2019, galeski_2021}, where claims of WTI properties close to DSM were reported.

 For the highest isoenergy in Fig.~\ref{fig:fs_main_text}\textit{d, f}, it is clear that the pair of Fermi pockets in STI and the single pocket in WTI are topologically distinct, where a topological phase transition should occur to turn one into another. Specifically, when applying adiabatic deformations, the separation between the pair of Fermi pockets, along the $Y - \Gamma - Y$ path, decreases to zero. At this point, they merge and form a single pocket due to the gap closing in the band structure. Later on, adiabatic deformations can further shape the elongated ellipsoid in WTI and cause it to evolve as discussed above. Nevertheless, symmetric Fermi pockets exist even under extreme deformations in WTI, such as \textit{Config.} 14 in SM, persisting on the $k_x-k_z$ plane. 

As shown in Fig.~\ref{fig:fs_main_text} and discussed above, it is evident that the topological properties of ZrTe$_5$ can also be identified by its isosurfaces. Moreover, the distinction of isosurfaces in STI and WTI can be experimentally detected. Thus, isosurfaces are more suitable than band structure patterns for identifying the corresponding topological phase of a ZrTe$_5$ sample.

%Distinguishing features between STI and WTI, as discussed previously in band structure, are much evident in their isosurfaces. Despite the equivalency of the chosen \textit{Configs.}, STI manifests dual symmetric Fermi pockets along the $k_y$-axis (see Fig.~\ref{fig:fs_main_text}a), while WTI showcases a single ellipsoidal pocket (similar to Fig.~\ref{fig:fs_main_text}b, c).. Even though the band structure in Fig.~\ref{fig:bands_reduced} implies dual Fermi pocket in the WTI regime for an augmented stacking distance, they emerge in a plane perpendicular to the $k_y$-axis. Yet, such extreme conditions may not warrant significant research attention for this material, given that the valence band maximum and conduction band minimum are considerably distant from the $\Gamma$ point, in contrast to the surface Dirac cone, which remains localized at this point. Beware that the difference originates from the topology of the system, making it impossible to adiabatically transform from one state to another.

%Finally, it is observable that certain Fermi surfaces exhibit flat extremities, a consequence of technical limitations. A more refined mesh in momentum space during simulations promises smoother surfaces.

%%%%%%%%%%%%%%%%%%%%%%%%%%%%%%%%%%%%%%%%%%%%%%%%%%%%%%%%%%%%%%%%%%%%
\subsection{Shubnikov-de Haas oscillations}
\label{subsec:qo}

Shubnikov-de Haas (SdH) effect is a typical quantum oscillation experiment widely performed in the studies of ZrTe$_5$~\cite{kamm_1985, izumi_1987,  wang_2018, tang_2019, galeski_2021, zhu_2022, kamm_1985}. We further calculated the quantum oscillations from our computed isosurfaces to compare with the experimental results. The results from Sec.~\ref{subsec:fs} indicate that extremal cross-sectional areas appear for $\vb{B} \parallel \vb{k}_y$, $\vb{B} \parallel \vb{k}_z$, and $\vb{B} \parallel \vb{k}_x$. Furthermore, symmetries of the isosurfaces show that the best way to find them is by rotating $\vb{B}$-field along the planes formed by the orthogonal wave vector basis $\vb{k}$, in the spherical coordinates.

\begin{figure*}[tb!]
	\centering
	\includegraphics[width=0.3\linewidth, height = 5.3cm]{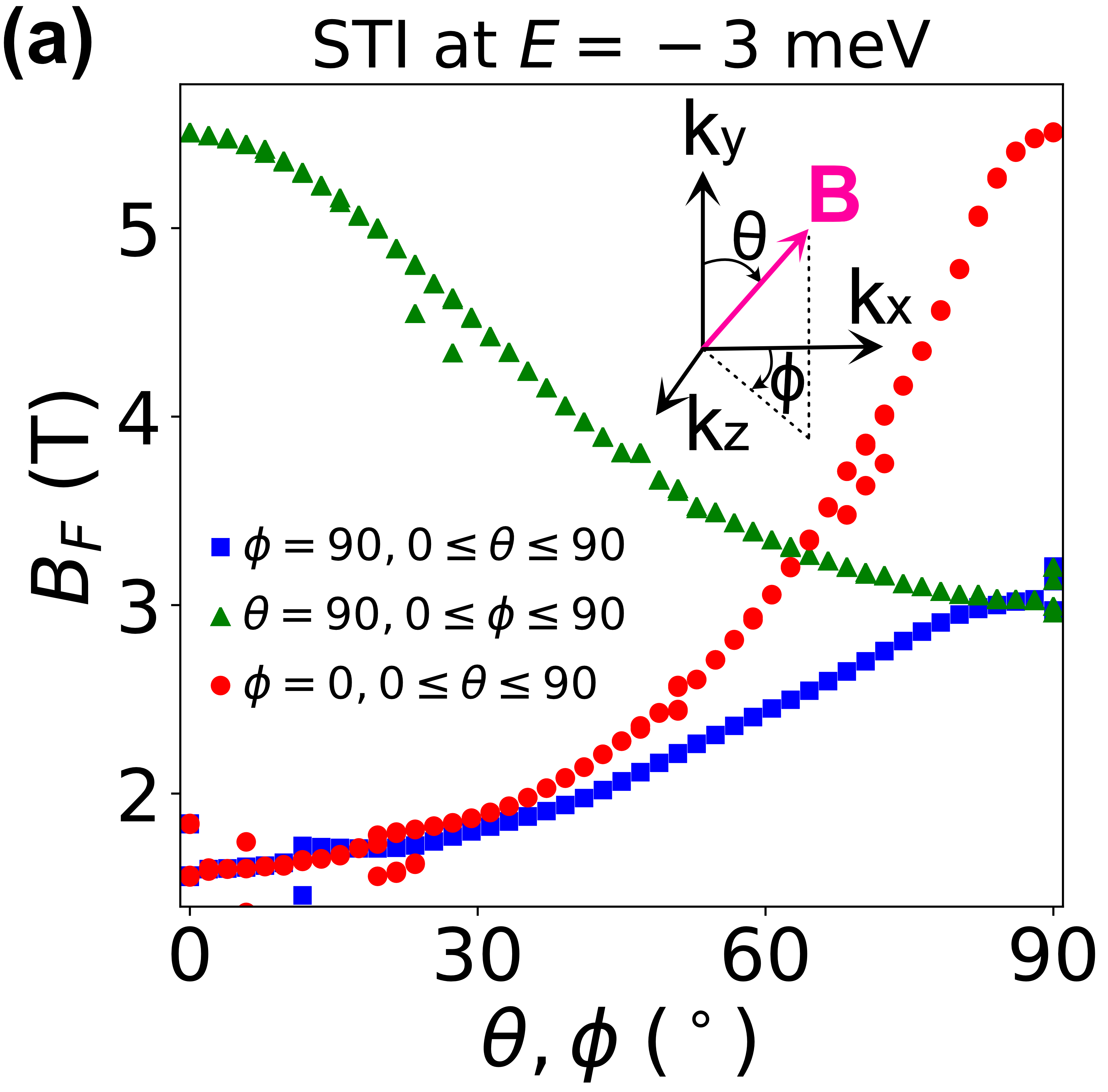}
	\includegraphics[width=0.3\linewidth, height = 5.3cm]{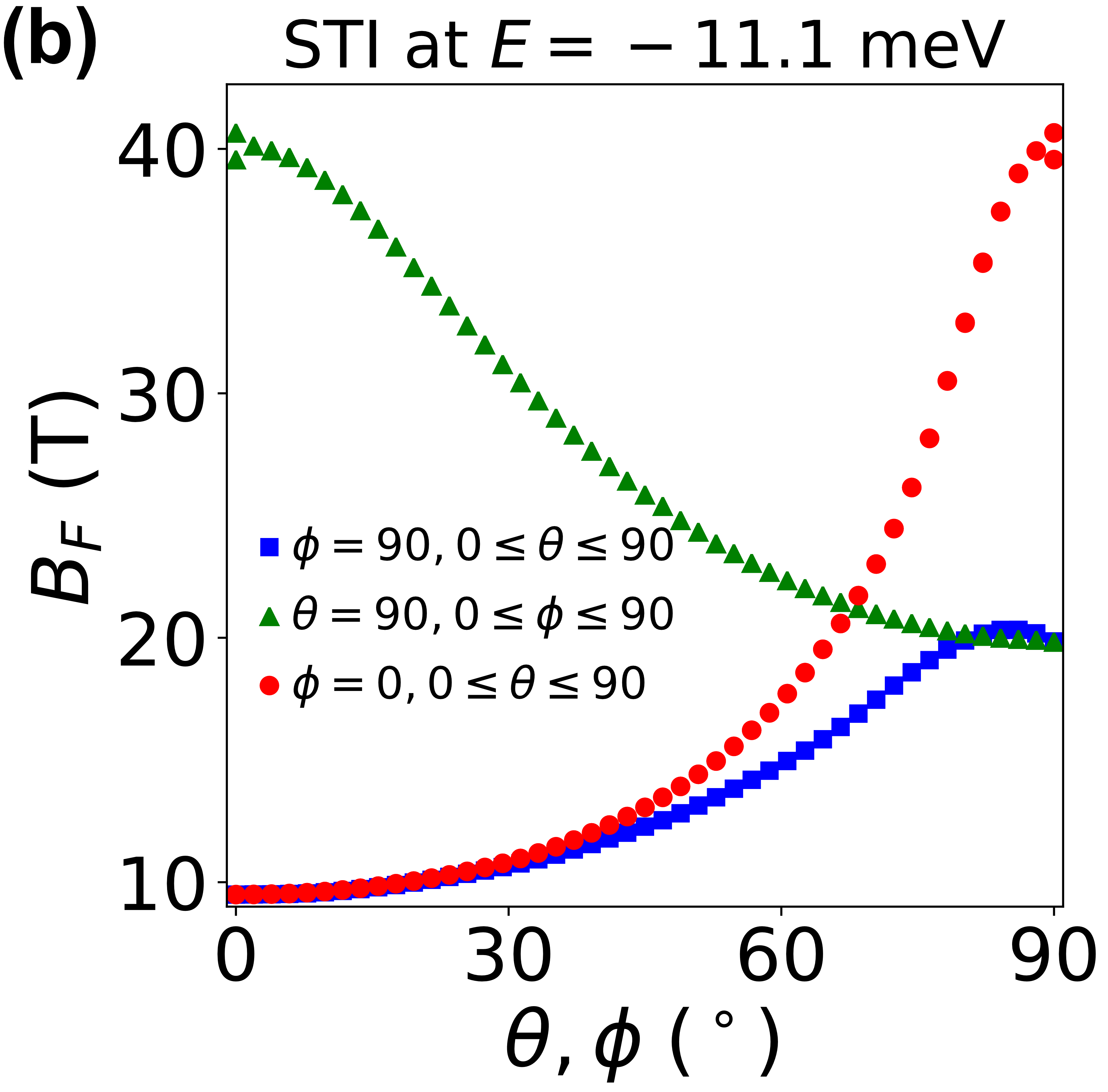}
	\includegraphics[width=0.3\linewidth, height = 5.3cm]{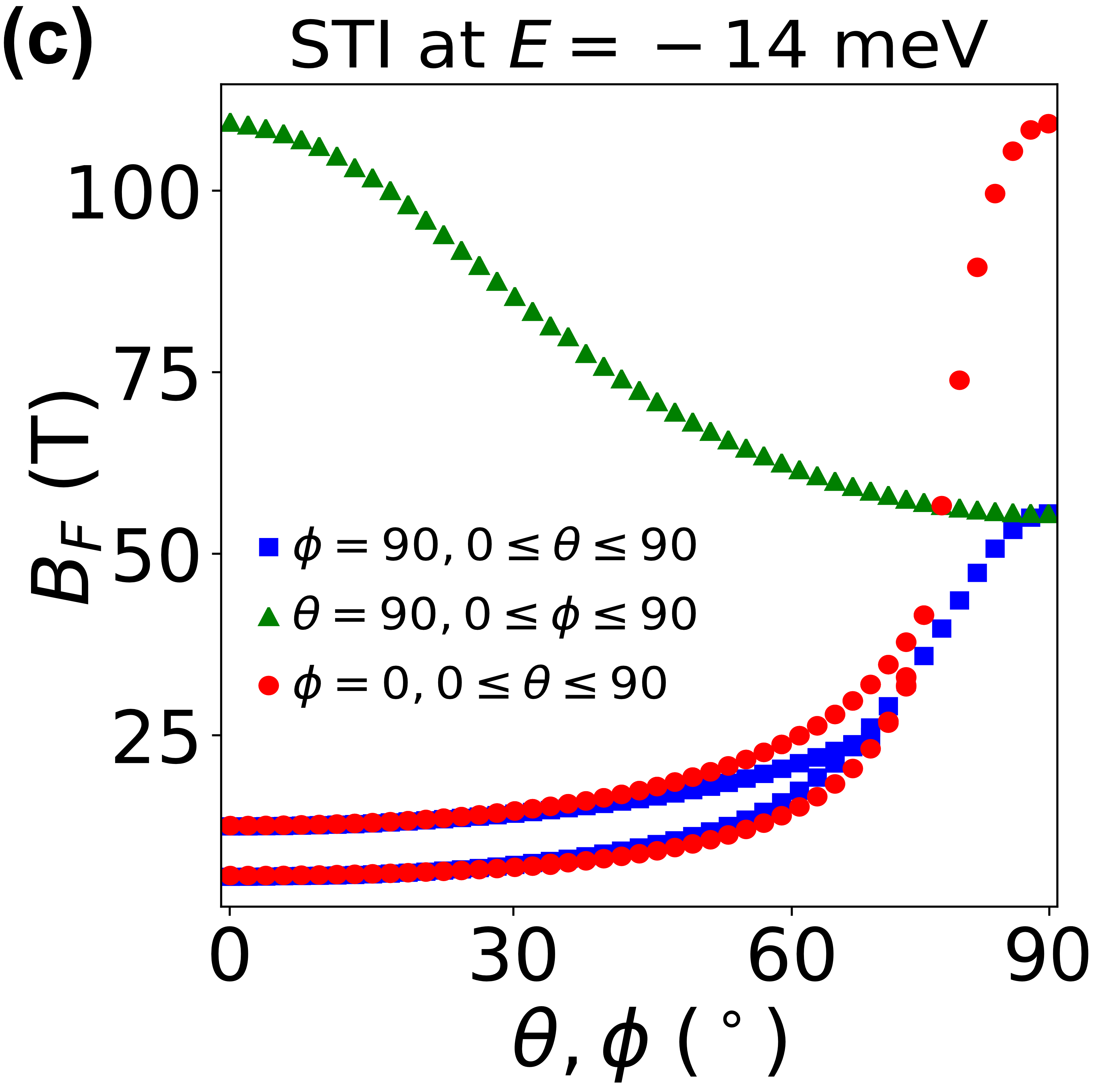}
	\includegraphics[width=0.3\linewidth, height = 5.3cm]{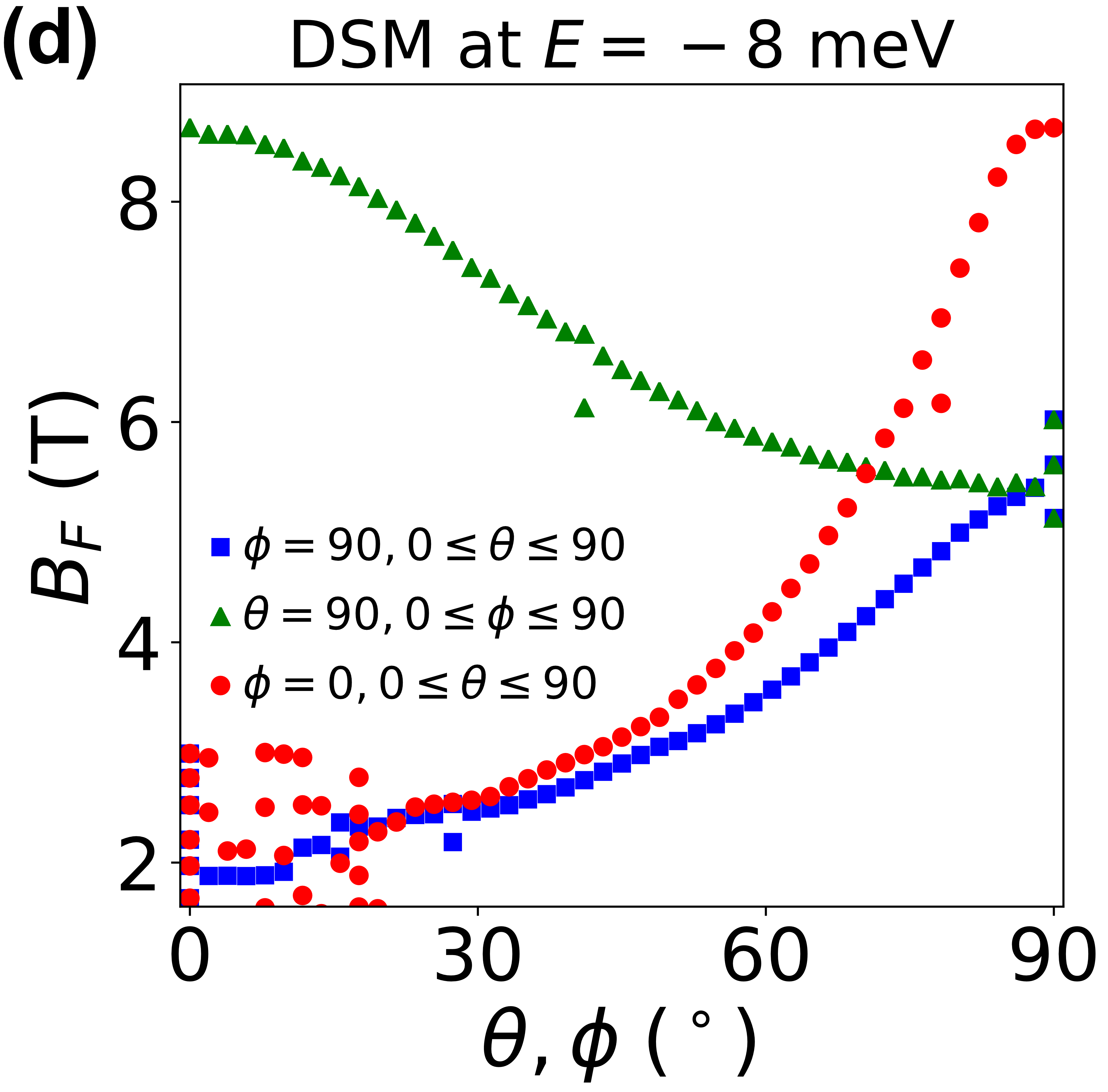}
	\includegraphics[width=0.3\linewidth, height = 5.3cm]{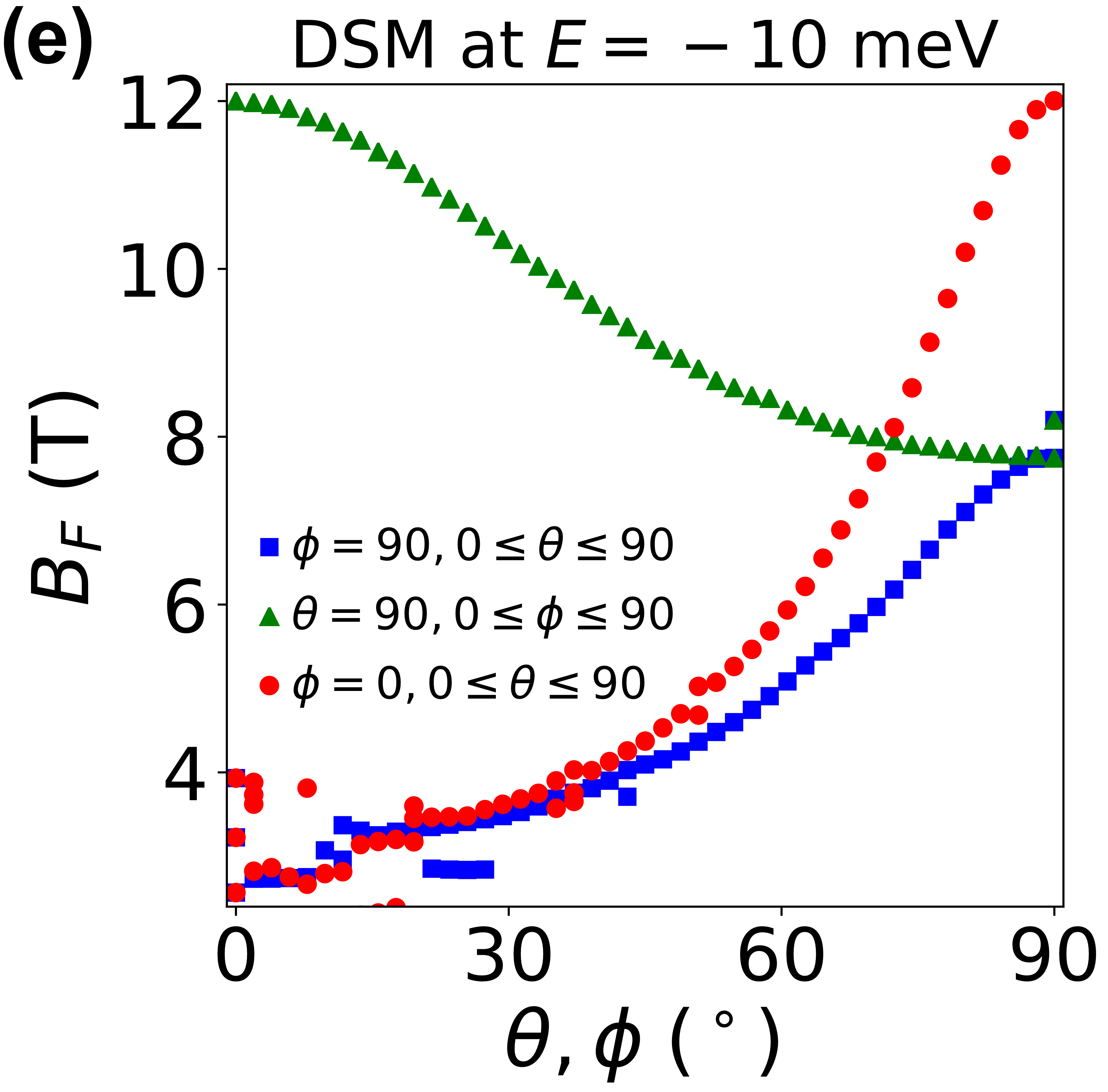}
	\includegraphics[width=0.3\linewidth, height = 5.3cm]{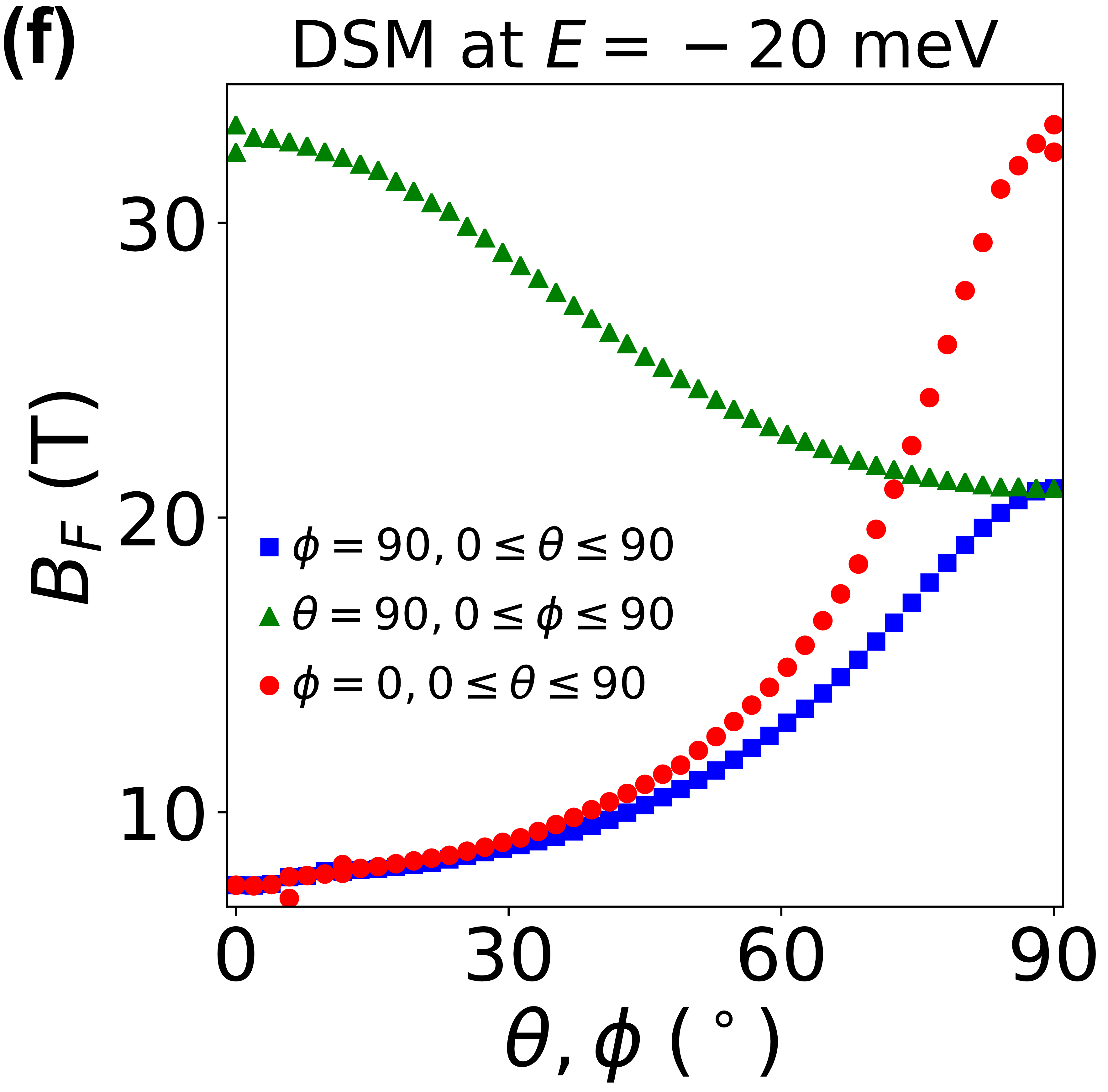}
	\includegraphics[width=0.3\linewidth, height = 5.3cm]{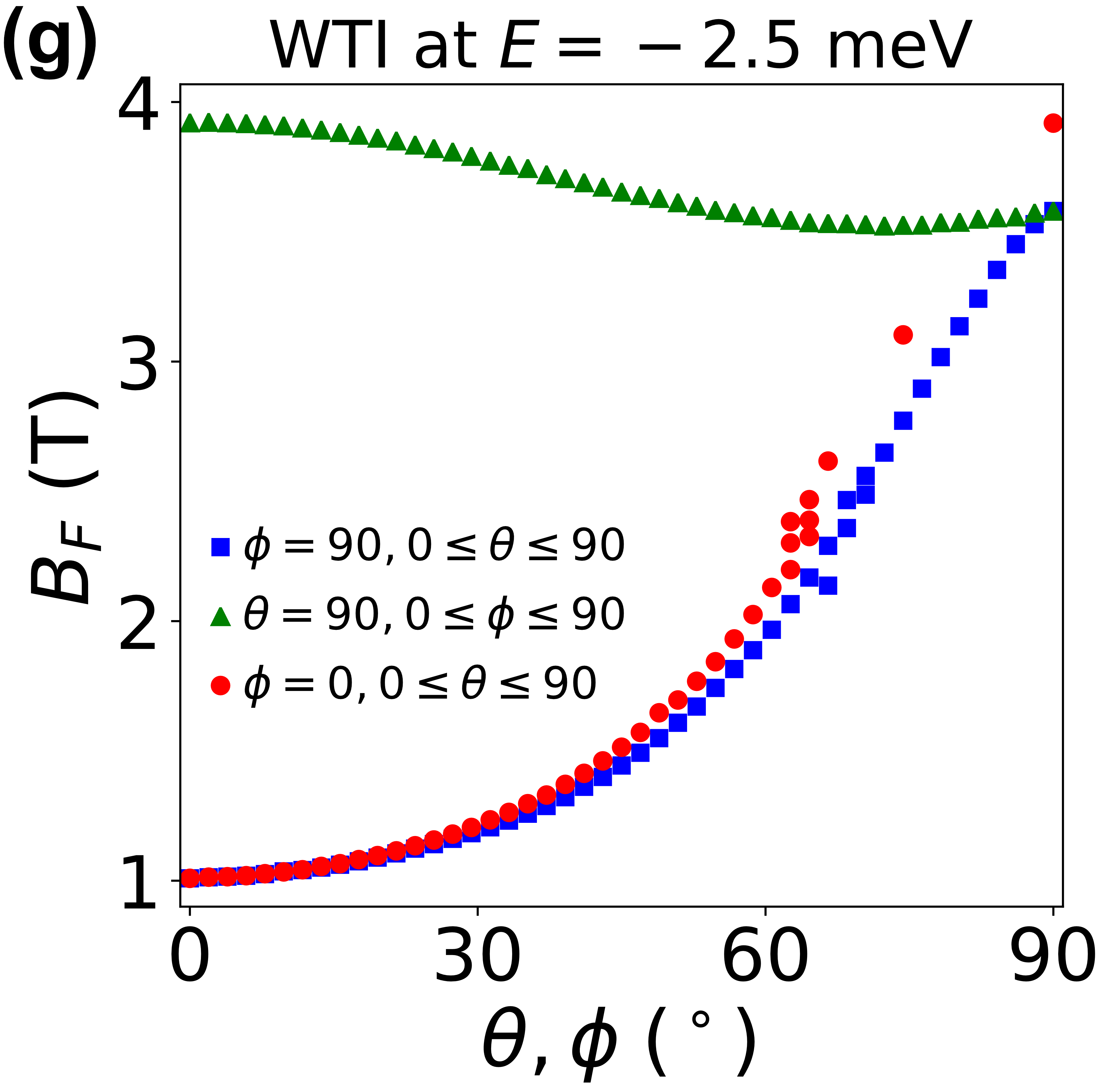}
	\includegraphics[width=0.3\linewidth, height = 5.3cm]{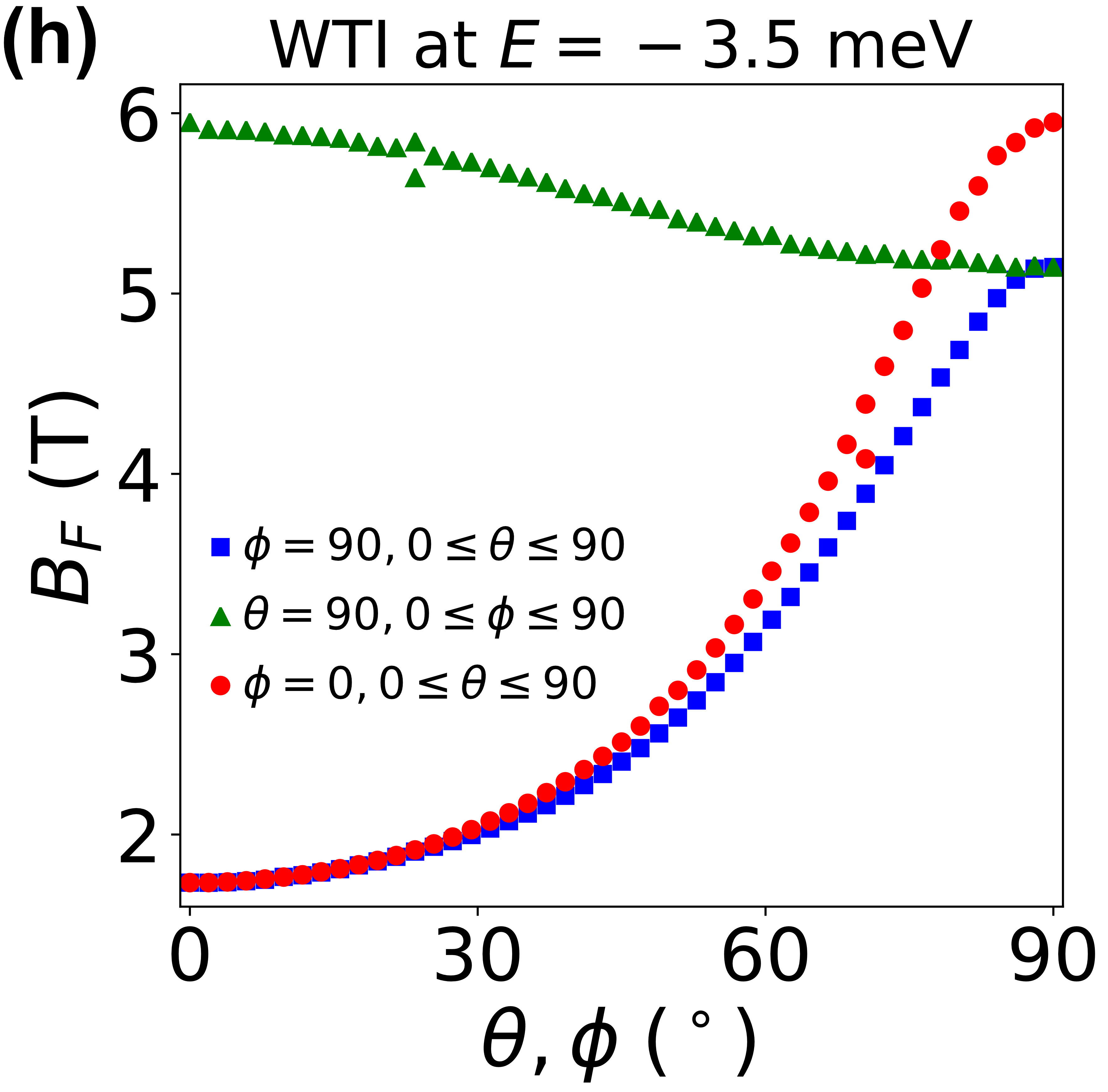}
	\includegraphics[width=0.3\linewidth, height = 5.3cm]{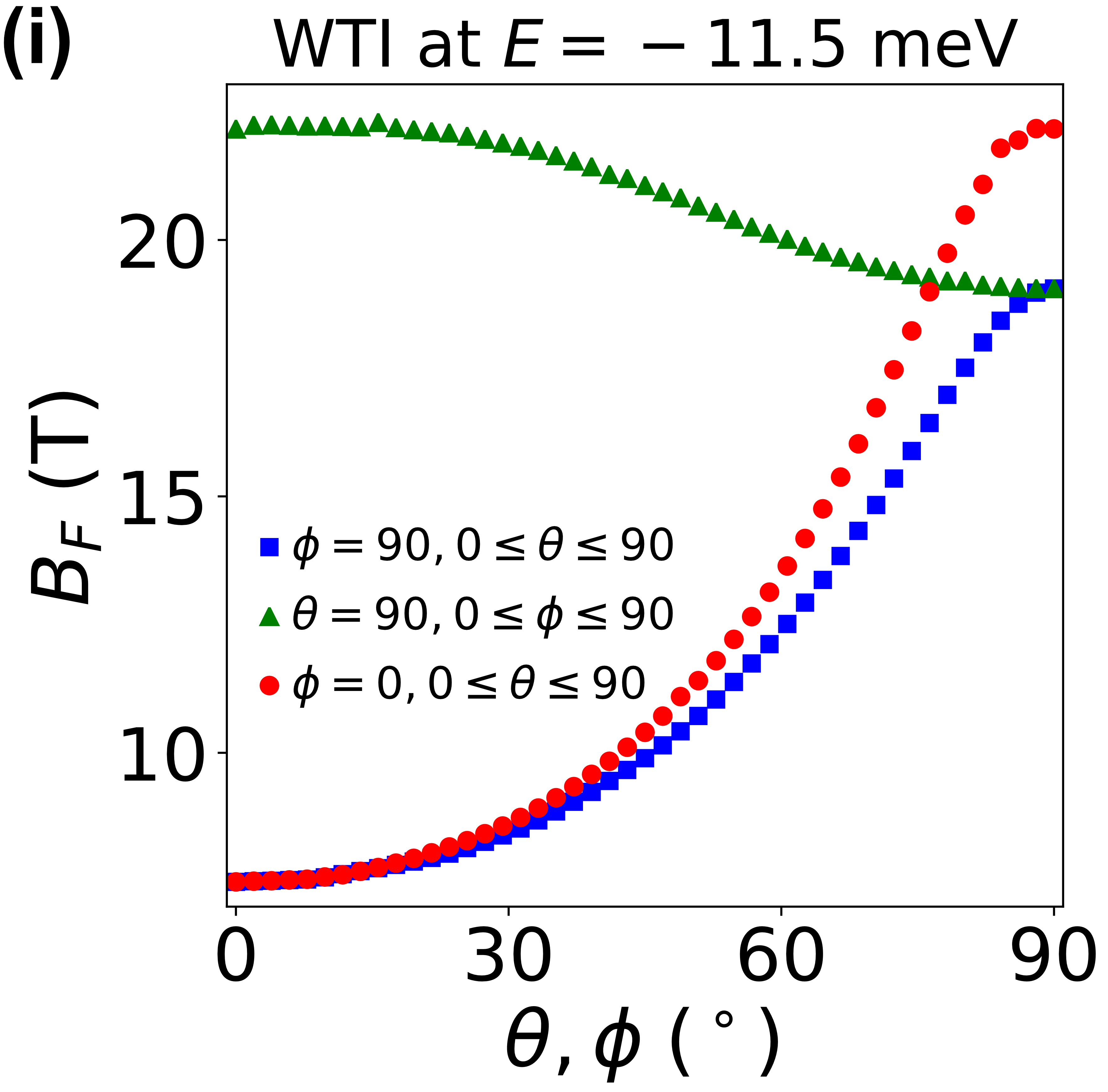}
\caption{Calculated SdH oscillations for different orientations of $\vb{B}$-field, for STI (\textit{Config.} $5$), DSM (\textit{Config.} $9$) and WTI (\textit{Config.} $12$) with different isosurfaces referring to the reference energy $E_{VBM} = 0$. Following the coordinate scheme in {(\textbf{a})}, the blue (\textit{i}.\textit{e}., square), green (\textit{i}.\textit{e}., triangle), and red (\textit{i}.\textit{e}., dot) values of $B_F$ are the results when rotating the magnetic field in the $k_y - k_z$,  $k_x - k_z$  and $k_x - k_y$ planes, respectively.}
	\label{fig:qo_frequencies}
\end{figure*}

Figure~\ref{fig:qo_frequencies} displays SdH frequencies, $B_F$, for different rotation angles of the applied $\vb{B}$-field, for STI (\textit{Config.} $5$), DSM (\textit{Config.} $9$) and WTI (\textit{Config.} $12$), calculated for different isosurfaces in the valence band. Following the coordinate scheme in Fig.~\ref{fig:qo_frequencies}\textit{a}, the blue (square), green (triangle) and red (dotted) values of $B_F$ are the results when rotating the $\vb{B}$-field in the $k_y - k_z$,  $k_x - k_z$  and $k_x - k_y$ planes, respectively. In general, the dispersed points that appear in many plots are caused by the small size of our isosurfaces, which are computationally costly to calculate in high resolution for smoothness. However, one can still interpolate the correct values based on the curves. The frequencies are higher for lower binding energies due to the increased size of the corresponding isosurfaces. In STI, Fig.~\ref{fig:qo_frequencies}\textit{a, b} only show a single curve for all directions of the applied magnetic field. This fact implies that the pair of symmetric Fermi pockets are not distinguishable in quantum oscillations. Nevertheless, one can still determine the number of symmetric Fermi pockets by studying the degeneracy in the quantum oscillation if the carrier doping can be accurately quantified, for example, by measuring the Hall effect or through field effect doping via a quantified condenser.

The double blue and red curves in Fig.~\ref{fig:qo_frequencies}\textit{c} imply two different extremal crossectional areas on the isosurface. This phenomenon happens mainly for isosurfaces in the STI phase because, along each magnetic field orientation, cutting a 3D ellipsoid can only form a single 2D plane, showing a single SdH frequency. This is exemplified by single curves in Fig.~\ref{fig:qo_frequencies}\textit{d-f} for DSM and in Fig.~\ref{fig:qo_frequencies}\textit{g-i} for WTI. The exception occurs for extreme deformations in WTI, where there will be more than one green curve according to our coordinates. Thus, having dual blue and red curves here is the direct proof of the STI phase, if experimentally observed. Comparing DSM and WTI, the blue and red curves are much closer for the latter phase because the ellipsoids are more uniform. This gives an insight into how close this WTI phase is concerning DSM. An extreme situation would be that the red curve becomes lower than the blue one, which means a flattened 3D ellipsoid. However, such a shape has not been reported yet. Finally, the energy difference of $1$ meV in Fig.~\ref{fig:qo_frequencies}\textit{g}, \textit{h} causes an increase of $\sim 2 \, T$ in $B_F$ for the highest frequency branch. This variation implies a highly sensitive dependence of SdH on the doping of the samples.

%with isosurface $E = -11 meV$: $\vb{B}$-field \textit{(a)} starts from $\vb{B} \parallel \vb{k}_x$ and ends up  $\vb{B} \parallel \vb{k}_z$, \textit{(b)} starts from $\vb{B} \parallel \vb{k}_y$ and ends up  $\vb{B} \parallel \vb{k}_x$, and \textit{(c)} starts from $\vb{B} \parallel \vb{k}_y$ and ends up  $\vb{B} \parallel \vb{k}_z$. Note that all the remaining results are for different isosurfaces and phases are similar, as detailed in S.~\ref{app:qo}. Firstly, figures imply a single isosurface, even though we know beforehand that the STI phase has dual Fermi pocket. Therefore, experimental SdH measurement may not reveal this phenomenon. Next, the lowest and highest values of $B_F$ belong to the extremal cross-sectional areas that we are looking for. These quantities, for the smallest possible isosurface of STI, DSM and WTI phases with valid SdH results according to the accuracy of our calculations, are listed in Tab.~\ref{tab:qo_main_text} along with the associated cyclotron mass. 

%Refs.~\cite{tang_2019, galeski_2021} The 3D ellipsoidal isosurface outcomes are consistent, and are in agreement with empirical outcomes presented in Refs.~\cite{tang_2019, galeski_2021}.

\begin{table*}[tb!]
	\centering
	\begin{tabular}{|c|c|c|c|c|}
		\hline
		\multicolumn{5}{|c|}{\textbf{Our Calculated Results}} \\  \hline
		Phase & $E \; (\mathrm{meV})$ & $\mathbf{B}$ direction & $B_F \; (\mathrm{T})$ & $m_{CR} \; (\mathrm{m_e})$ \\ \hline
		%		& & $\mathbf{B} \parallel \mathbf{k}_y$ & 1.598 & 0.104 \\
		%		STI & -29 & $\mathbf{B} \parallel \mathbf{k}_z$ & 3.028 & 0.191 \\
		%		& & $\mathbf{B} \parallel \mathbf{k}_x$ & 5.508 & 0.321 \\ \hline \hline
		& & $\mathbf{B} \parallel \mathbf{k}_y$ & 1.879 & 0.047 \\
		DSM & -8 & $\mathbf{B} \parallel \mathbf{k}_z$ & 5.403 & 0.103 \\
		& & $\mathbf{B} \parallel \mathbf{k}_x$ & 8.673 & 0.175 \\ \hline \hline
		& & $\mathbf{B} \parallel \mathbf{k}_y$ & 1.051 & 0.075 \\
		WTI & -2.5 & $\mathbf{B} \parallel \mathbf{k}_z$ & 3.580 & 0.187 \\
		& & $\mathbf{B} \parallel \mathbf{k}_x$ & 3.970 & 0.215 \\ \hline
	\end{tabular}
	\hspace{2cm}
	\begin{tabular}{|c|cc|cc|}
		\hline
     	& \multicolumn{2}{c|}{\textbf{Tang \textit{et al.}}}                & \multicolumn{2}{c|}{\textbf{Galeski \textit{et al.}}}             \\ \hline
 $\vb{B}$ direction         & \multicolumn{1}{c|}{$B_F (\mathrm{T})$} & $m_{CR} \; (\mathrm{m_e})$ & \multicolumn{1}{c|}{$B_F (\mathrm{T})$} & $m_{CR} \; (\mathrm{m_e})$ \\ \hline
 $\vb{B} \parallel \vb{k}_y $ & \multicolumn{1}{c|}{1.32}      & 0.016          & \multicolumn{1}{c|}{1.1}       &                \\ $\vb{B} \parallel \vb{k}_z $ & \multicolumn{1}{c|}{9.9}       & 0.091          & \multicolumn{1}{c|}{12.3}      &                \\ $\vb{B} \parallel \vb{k}_x $ & \multicolumn{1}{c|}{16.6}      & 0.23           & \multicolumn{1}{c|}{16.7}      &                \\ \hline \hline
 $\vb{B} \parallel \vb{k}_y $ & \multicolumn{1}{c|}{1.18}      & 0.016          & \multicolumn{1}{c|}{1.2}       & 0.04           \\
 $\vb{B} \parallel \vb{k}_z $ & \multicolumn{1}{c|}{9.2}       & 0.12           & \multicolumn{1}{c|}{13.9}      & 0.36           \\ $\vb{B} \parallel \vb{k}_x $ & \multicolumn{1}{c|}{15.7}      & 0.344          & \multicolumn{1}{c|}{15.2}      & 0.68           \\ \hline
	\end{tabular}
	\caption{(\textbf{\textit{Left}}) Calculated extremal SdH frequencies and cyclotron masses of DSM (\textit{i.e.}, \textit{Config.} $9$), and WTI (\textit{i.e.}, \textit{Config.} $12$) for isosurfaces close to the corresponding VBM. (\textbf{\textit{Right}}) Experimental SdH frequencies and cyclotron masses from Tang \textit{et al.}~\cite{tang_2019} and Galeski \textit{et al.}~\cite{galeski_2021}. }
	\label{tab:qo_main_text}
\end{table*}

To compare with experimental results, we consider our extremal values of $B_F$ for the smallest size of isosurface possible in STI and WTI. Table~\ref{tab:qo_main_text} lists our results \textit{(left table)} to compare with measurements from Tang \textit{et al.}~\cite{tang_2019} and Galeski \textit{et al.}~\cite{galeski_2021} \textit{(right table)}. In general, the calculated $B_F$ for both $\mathbf{B} \parallel \mathbf{k}_z$ and $\mathbf{B} \parallel \mathbf{k}_x$ are smaller than the experimental ones. This means the 3D ellipsoid in our calculations is less elongated than in the measurements. One of the main factors that influence the shape of the ellipsoid is the interlayer separation, which also determines the distance between $\Gamma$ and $Y$ points in reciprocal space. Furthermore, our calculated $m_{CR}$ is bigger than the experimental ones for both $\mathbf{B} \parallel \mathbf{k}_y$ and $\mathbf{B} \parallel \mathbf{k}_z$, but smaller for $\mathbf{B} \parallel \mathbf{k}_x$. A possible explanation for these differences might come from the sensitive structural changes in different materials. In our \textit{ab initio} simulations, the relaxed structure is an ideal crystal in a vacuum. In contrast, real samples are synthesized under many different conditions resulting in distinct samples with slightly different structures, especially when processed in different research groups~\cite{tang_2019, galeski_2021}.

\section{Conclusions}
\label{sec:conclusions}

In conclusion, we have performed \textit{ab initio} DFT simulations of 3D bulk ZrTe$_5$ to track the topological phase transition and the concomitant Shubnikov-de Haas (SdH) oscillations. We have uncovered that isoenergetic surfaces in the reciprocal space capture the entire process of the topological phase transition. Tracing the variations in Fermi surfaces allows one to determine, in both theory and experiments, the corresponding topological phases without computing topological invariants or contributions from boundary states. To be more specific, isosurfaces in Dirac semimetal with a closed bulk gap always have an elongated 3D non-uniform ellipsoidal shape. Isosurfaces in WTI are mainly 3D non-uniform ellipsoids, persisting even at large structural deformations. Furthermore, these calculated ellipsoidal isosurfaces can compare with the results measured by experiments~\cite{tang_2019, galeski_2021}. In contrast, the isosurface for a strong topological insulator (STI) generally has a pair of symmetric Fermi pockets near the valence band maximum. When going gradually to lower isoenergies, corresponding to larger hole doping, the two Fermi pockets start to merge and evolve into a dumbbell shape. In the end, it turns into an elongated 3D non-uniform ellipsoid.

Moreover, we have unveiled that the calculated SdH oscillation frequencies can effectively distinguish STI between WTI phases. When rotating the magnetic field, the WTI always has a single SdH oscillation frequency for all orientations, even for large structural deformations. An isosurface with a dumbbell shape along $k_y$-axis in STI implies two oscillation frequencies for an interval of orientations. In the case of a pair of Fermi pockets along $k_y$-axis in STI, one can combine with the analysis on degeneracy to correctly determine the corresponding topological phase. The results also indicate whether a sample characterized in a WTI phase is close to DSM phase. Specifically, the ellipsoid in DSM is quite non-uniform, but the one in WTI becomes more uniform when it stays further away from DSM. Finally, we have compared our calculated SdH results with the experimental ones, and found that they are very sensitive to how the crystal lattice parameters and doping vary. 

We have successfully identified the entire process of topological phase transition in the energy band structures. Moreover, we have disclosed a pattern in isosufaces that allows one to distinguish between STI and WTI. In STI, the valence band shows two symmetric peaks along the $Y - \Gamma - Y $ path, which explains the pair of Fermi pockets mentioned above. Conversely, WTI mostly has a Gaussian-type maximum along this path, except for extreme structural deformations; the symmetric valence peaks appear along the $X - \Gamma - X $ and along the $A - \Gamma - A $ paths for extreme cases. This pattern can be used, for instance, to compare with angle-resolved photoemission spectroscopy (ARPES) measurements or to study a few layers of ZrTe$_5$. Finally, we have found that $\Delta c_2$-axis in the conventional cell is the key term responsible for the bulk gap closing, which was not clear from the prior theoretical and experimental studies~\cite{weng_2014, manzoni_2016, fan_2017}. Overall, the studies of fermiology directly reveals different topological phases of bulk ZrTe$_5$; we believe that similar behavior could be observed in other topological systems.

\section*{\label{sec:Contributions}Data Availability}
The data that support the findings of this study will be openly available at DataVerseNL after the paper is accepted. 

\bibliographystyle{naturemag}
\bibliography{lib.bib}

%%%%%%%%%%%%%%%%%%%%%%%%%%%%%%%%%%%%%%%%%%%%%%%%%%%%%%%%%%%%%%%%%%%%
\begin{acknowledgments}         
	We thank K. Tenzin, H. Jafari, and E. Barts for the helpful discussions. The calculations were carried out on the Dutch national e-infrastructure with the support of SURF Cooperative (EINF-8334) and the Hábrók high-performance computing cluster of the University of Groningen. We acknowledge the research program “Materials for the Quantum Age” (QuMat) for financial support. This program (registration number 024.005.006) is part of the Gravitation program financed by the Dutch Ministry of Education, Culture and Science (OCW).
\end{acknowledgments}

\vspace{0.6cm}

\section*{\label{sec:Contributions}Author Contributions}
C.C.Y. performed DFT calculations and wrote the initial draft of the manuscript. All the authors analyzed the results and co-wrote the manuscript. J.S. supervised the project.

\section*{\label{sec:Contributions}Competing Interests}
The authors declare no competing interests.

\section*{\label{sec:Contributions}Additional Information}
Supplementary information. The online version contains supplementary material available at XXX.
	
% \FloatBarrier
% %%%%%%%%%%%%%%%%%%%%%%%%%%%%%%%%%%%%%%%%%%%%%%%%%%%%%%%%%%%%%%%%%%%%
% \bibliographystyle{naturemag}
% \bibliography{lib.bib}

%%%%%%%%%%%%%%%%%%%%%%%%%%%%%%%%%%%%%%%%%%%%%%%%%%%%%%%%%%%%%%%%%%%%
%%%%%%%%%% Merge with supplemental materials %%%%%%%%%%
%%%%%%%%%% Prefix a "S" to all equations, figures, tables and reset the counter %%%%%%%%%%
% \clearpage
% \setcounter{equation}{0}
% \setcounter{figure}{0}
% \setcounter{table}{0}
% \setcounter{page}{1}
% \makeatletter
% \renewcommand{\theequation}{S\arabic{equation}}
% \renewcommand{\thefigure}{S\arabic{figure}}
% \renewcommand \thesection{S\@arabic\c@section}
% \renewcommand\thetable{S\@arabic\c@table}
% \renewcommand{\bibnumfmt}[1]{[S#1]}
% \renewcommand{\citenumfont}[1]{S#1}
% \makeatother
% %%%%%%%%%% Prefix a "S" to all equations, figures, tables and reset the counter %%%%%%%%%%

% \FloatBarrier

% \input{supplementary.tex}

% \bibliographystyle{naturemag}
% \bibliography{lib.bib}

%%%%%%%%%%%%%%%%%%%%%%%%%%%%%%%%%%%%%%%%%%%%%%%%%%%%%%%%%%%%%%%%%%%%
\end{document}